\documentclass[twocolumn,english,aps,pra,showpacs]{revtex4}
\usepackage{amssymb}
\usepackage{amsmath}
\usepackage{colordvi}
\usepackage{amsthm}
\usepackage{epsfig}
\usepackage{graphicx}
\usepackage{subfigure}
\usepackage{color}

\def\avg(#1){\langle#1\rangle}

\def\be{\begin{equation}}
\def\ee{\end{equation}}
\def\bea{\begin{eqnarray}}
\def\eea{\end{eqnarray}}

\begin{document}

\title{Two-component polariton condensate in optical microcavity}
\author{Yong-Chang Zhang$^{\text{1,2}}$}
\author{Xiang-Fa Zhou$^{\text{1,2}}$}
\author{Guang-Can Guo$^{\text{1,2}}$}
\author{Xingxiang Zhou$^{\text{1,2}}$}
\email{xizhou@ustc.edu.cn}
\author{Han Pu$^{\text{3}}$}
\email{hpu@rice.edu}
\author{Zheng-Wei Zhou$^{\text{1,2}}$}
\email{zwzhou@ustc.edu.cn}
\address{$^{\text{1}}$Key Laboratory of Quantum Information,
University of Science and \\ Technology of China, Hefei, Anhui
230026, P. R. China\\ $^{\text{2}}$Synergetic Innovation Center
of Quantum Information and Quantum Physics, University of Science
and Technology of China, Hefei, Anhui 230026, China\\ $^{\text{3}}$Department of Physics and Astronomy, and Rice Quantum Institute, Rice University, Houston, TX 77005, USA}

\begin{abstract}
We present a scheme for engineering the extended two-component Bose-Hubbard model using polariton condensate supported by optical microcavity. Compared to the usual two-component Bose-Hubbard model with only Kerr nonlinearity, our model includes a nonlinear tunneling term which depends on the number difference of the particle in the two modes. In the mean field treatment, this model is an analog to a nonrigid pendulum with a variable pendulum length whose sign can be also changed. We study the dynamic and ground state properties of this model and show that there exists a first-order phase transition as the strength of the nonlinear tunneling rate is varied. Furthermore, we propose a scheme to obtain the polariton condensate wave function.
\end{abstract}
\maketitle

\section{Introduction}
As a kind of new state of matter, Bose-Einstein condensation (BEC) in a dilute gas of trapped atoms is able to exhibit quantum phenomena on macroscopic scales \cite{BEC1,BEC2,BEC3,BEC4}. Different from pure single-particle behaviour, the interparticle interaction between condensate atoms gives rise to many intriguing nonlinear phenomena. One important example is the dynamics of a condensate trapped in a double-well potential. In this system, the condensate undergoes either Josephson oscillation where the population oscillates sinusoidally between the two wells, or self trapping in which most atoms remain trapped in one of the wells. This has been theoretically predicted \cite{Shenoy} and experimentally demonstrated \cite{Ketterle,Oberthaler,Steinhauer}. Recently, in such a double-well model, the quantum phase transition (QPT) and dynamics induced by atom-pair tunneling had also been theoretically investigated \cite{Liang,Fu}.


In recent years, BEC of microcavity polaritons has been experimentally demonstrated \cite{Deng,Snoke,Shelykh}. As a kind of bosonic quasiparticle, the polariton represents the excitation of the eigenmodes of the light-matter system inside the microcavity, which can be manipulated and generated by the external laser field. The interaction between the polaritons can take the form of Kerr nonlinearity, which occurs if atoms with a specific level structure interact with light \cite{Werner}. Such kind of controllable and strong Kerr nonlinearity can be used to simulate strongly correlated many-body models in photon-coupled microcavity arrays \cite{Plenio}.

In this paper we present a scheme for simulating the tunneling between two polariton condensates in the microcavity system. Here, compared with the case of BEC in a double-well potential, two spatially localized bosonic modes are replaced with two different modes of polaritons in the same cavity. Under proper arrangement, the effective nonlinear tunneling between the two  polariton condensates can be easily induced and controlled by external fields. In our work, we derive a two-mode model that describes the system. A salient feature is that the tunneling rate between the two modes takes a nonlinear form that depends on the difference of the population in the modes. Under the mean-field approximation, such nonlinear Hamiltonian will be reduced to a new nonrigid pendulum model, in which not only the magnitude, but also the sign of the pendulum length can be changed. This nonlinear tunneling directly leads to the emergence of new dynamic phases in addition to Josephson oscillation and self-trapping.

Our paper is organized as follows: in Sec. II, we give out a physical realization for the extended two component-Bose Hubbard model (ETC-BHM) in the microcavity polariton system. In the regime of polariton condensate, we investigate the semiclassical behavior and dynamical properties of this model in Sec. III and IV respectively. Section V displays the ground state properties of ETC-BHM. Here, we find that there is a first-order quantum phase transition in the proper coefficient regime. In Sec. VI, we present a scheme to extract the information on polariton condensate wave function. Finally, we conclude in Sec. VII.

\section{Model Hamiltonian}
\begin{figure}
\includegraphics[width=7.cm]{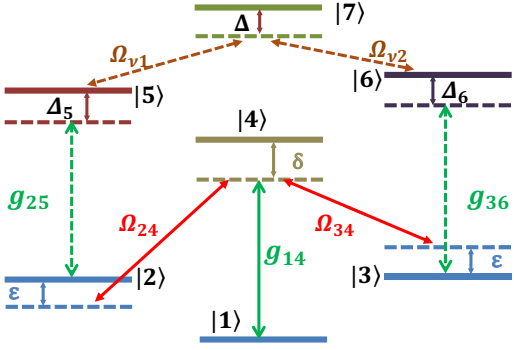}
\caption{(Colour online) Schematic diagram for the energy levels of the ensemble of the atoms trapped in the microcavity. The cavity mode with frequency $\omega_c$ drives the transitions: $\vert 1{\rangle}\leftrightarrow \vert 4{\rangle}$, $\vert 2{\rangle}\leftrightarrow \vert 5{\rangle}$, and $\vert 3{\rangle}\leftrightarrow \vert 6{\rangle}$ with strengths $g_{14}$, $g_{25}$, $g_{36}$ {and detunings $\delta$, $\Delta_5$, $\Delta_6$}, respectively. The driving laser fields $\Omega_{24}$ and $\Omega_{34}$ induce the transitions: $\vert 2{\rangle}\leftrightarrow \vert 4{\rangle}$ and $\vert 3{\rangle}\leftrightarrow \vert 4{\rangle}$ {with two-photon detuning $\varepsilon$}. The microwave fields $\Omega_{\nu1}$ and $\Omega_{\nu2}$ induce the transitions: $\vert 5{\rangle}\leftrightarrow \vert 7{\rangle}$ and $\vert 6{\rangle}\leftrightarrow \vert 7{\rangle}$ {with detuning $\Delta$}.}
\label{level}
\end{figure}

We consider an ensemble of $N_a$ bosonic atoms inside a single-mode optical cavity with the frequency $\omega_c$. As schematically shown in Fig.~\ref{level}, each atom has seven relevant hyperfine energy levels, three of which (states $|1 \rangle$, $|2 \rangle$, and $|3 \rangle$) belong to the electronic ground  manifold, and the other four (states $|4 \rangle$, $|5 \rangle$, $|6 \rangle$, and $|7 \rangle$) belong to the electronic excited manifold. The ground states  $|1 \rangle$, $|2 \rangle$, and $|3 \rangle$ are dipole coupled to the excited states $|4 \rangle$, $|5 \rangle$, and $|6 \rangle$, respectively, by the cavity field, with corresponding coupling strengths $g_{14}$, $g_{25}$ and $g_{36}$. States $|2 \rangle$ and $|3 \rangle$ are coupled to $|4 \rangle$ by external {laser fields} with coupling strengths $\Omega_{24}$ and $\Omega_{34}$, respectively. Finally, within the excited manifold, states $|5 \rangle$ and $|6 \rangle$ are coupled to $|7 \rangle$ by microwave fields with corresponding coupling strengths $\Omega_{\nu 1}$ and $\Omega_{\nu 2}$. The total Hamiltonian that describes this system can be written as:
\begin{equation}
\begin{split}
H\!=\!&\ \omega_c \left(a^{\dagger}a+\frac{1}{2} \right)+\sum_{j=1}^{N_a} \left\{ \sum_{i=1}^7 \omega_i \vert i{\rangle}_{jj} \langle i\vert \right.
\\&+[\Omega_{24}\vert 2{\rangle}_{jj} \langle 4\vert \cos (\omega_{L1}t)
+\Omega_{34}\vert 3{\rangle}_{jj}\langle 4\vert \cos (\omega_{L2}t)+h.c.]
\\&+[\Omega_{\nu1}\vert 5{\rangle}_{jj} \langle 7\vert \cos (\omega_{\nu1}t)
+\Omega_{\nu2}\vert 6{\rangle}_{jj}\langle 7\vert \cos (\omega_{\nu2}t)+h.c.]
\\&+ \left. [(g_{14}\vert 1{\rangle}_{jj}\langle 4\vert + g_{25}\vert 2{\rangle}_{jj}\langle 5\vert +g_{36}\vert 3{\rangle}_{jj}\langle 6\vert)a^{\dagger}+h.c.] \right\}.
\end{split}
\label{eq:ne1}
\end{equation}
Here, $a$ and $a^\dag$ are annihilation and creation operators for the cavity field, $\omega_{L_i}$ and $\omega_{\nu_i}$ ($i=1,2$) refer to the frequencies of the external laser fields and the microwave fields, respectively.

Under the situation that the fields are weak, the atoms mostly occupy the ground level $|1 \rangle$. In Appendix \ref{app1}, we will show that under proper conditions, the system can described by an effective two-mode Hamiltonian:
\begin{equation}
\begin{split}
\mathcal{H}_{\rm eff}&\approx \frac{V_1}{2}P^{\dagger2}_1 P^2_1 +\frac{V_2}{2}P^{\dagger2}_2 P^2_2 +UP^{\dagger}_1 P_1 P^{\dagger}_2 P_2 \\
& \quad  + (N-1)T^+[P^{\dagger}_1 P_2 +P^{\dagger}_2 P_1]\\
&\quad +T^- [P^{\dagger}_1(P^{\dagger}_1 P_1 -P^{\dagger}_2 P_2)P_2 +P^{\dagger}_2(P^{\dagger}_1 P_1 -P^{\dagger}_2 P_2)P_1]\,,
\end{split}
\label{eq:ne6}
\end{equation}
where {$N=\langle P^\dagger_1P_1 +P^\dagger_2P_2 \rangle$ is the total number of the polaritons (Note that the total number operator $P^\dagger_1P_1 +P^\dagger_2P_2$ commutes with the effective Hamiltonian $\mathcal{H}_{\rm eff}$, so we could replace it with its expectation value $N$ which is a constant)}, and in terms of the cavity photon operator and the collective atomic operators
\begin{equation}
S_{1i}=\frac{1}{\sqrt{N_a}}\sum_{j=1}^{N_a}\vert 1\rangle_{jj}\langle i\vert,
\label{eq:ne2}
\end{equation}
the creation operators for the two polariton modes are defined as
\begin{equation}
P^{\dagger}_{1(2)}\doteq \frac{1}{2} \left(\frac{g}{\omega}\pm 1\right) S^{\dagger}_{12}+\frac{1}{2} \left(\frac{g}{\omega} \mp 1 \right) S^{\dagger}_{13}-\frac{\Omega}{\sqrt{2}\omega}a^{\dagger},
\end{equation}
where $\omega =\sqrt{g^2 +\Omega^2}$, and for simplicity we have taken $\Omega_{24}=\Omega_{34} \equiv \sqrt{2} \Omega$ and $g=\sqrt{N_a} g_{14}$. In the weak-field limit, the collective atomic operators obey the bosonic commutation relation: $[S_{1i}\ ,\ S^{\dagger}_{1j}]=\delta_{ij}$, $[S_{1i}\ ,\ S_{1j}]=0$, and $[S^{\dagger}_{1i}\ ,\ S^{\dagger}_{1j}]=0$, from which one can readily show that the polariton operators also satisfy $ [P_i, P_j^\dag]=\delta_{ij}$ ($i,j=1,2$).

We name Hamiltonian (\ref{eq:ne6}) the ETC-BHM. In the ETC-BHM, the first three terms depict the Kerr nonlinearity. The terms proportional to $T^+$ represents the linear ``tunneling" or the conversion between the two modes with a tunneling rate given by $(N-1)T^+$. The terms proportional to $T^-$ can be regarded as a nonlinear tunneling term whose effect is to convert one mode into the other howevever with a conversion rate proportional to the population difference between the two modes. In Appendix~\ref{app1}, we provide an intuitive picture to explain the physical origin of various terms. It is the distinct effects induced by the nonlinear tunneling term that we will pay particular attention to.

\section{Semiclassical Hamiltonian}
We confine our discussion on the ETC-BHM in the regime of polariton condensate. Under this situation, the mean-field treatment is suitable. To this end, we replace the operators in Hamiltonian (\ref{eq:ne6}) by their respective expectation values: $P_\alpha \rightarrow \langle P_\alpha \rangle=\sqrt{N_\alpha(t)}e^{-i\theta_\alpha(t)}$ ($\alpha=1,2$) with $N_\alpha(t)$ and $\theta_\alpha(t)$ being the $\alpha$th polariton condensate's occupation number and phase, respectively \cite{Shenoy,Liu,Pu}. The equations of motion can be easily derived as:
\begin{eqnarray}
\frac{d \xi}{d t} &=& 2N(T^+ +T^-\xi)\sqrt{1-\xi^2}\sin\theta \,,
\label{eq:ne7} \\
\frac{d \theta}{d t} &=&\frac{N(V_1-V_2)}{2}+\frac{N(V_1+V_2-2U)}{2}\xi \nonumber \\
&& -2N\frac{T^+\xi-T^-(1-2\xi^2)}{\sqrt{1-\xi^2}}\cos{\theta} \,.
\label{eq:ne8}
\end{eqnarray}
Here, $\xi=\frac{N_1(t)-N_2(t)}{N}$ is the population difference and $\theta=\theta_2(t)-\theta_1(t)$ is the relative phase between the two condensates. In classical mechanics, the above two ordinary differential equations can be recast into the canonical form by the Hamiltonian equation:
\begin{equation}
\frac{d \xi}{d t}=-\frac{\partial H_c}{\partial \theta}\,, \qquad\qquad \frac{d \theta}{d t}=\frac{\partial H_c}{\partial \xi},
\label{eq:ne9}
\end{equation}
where the corresponding semiclassical Hamiltonian is:
\begin{eqnarray}
H_c &=& \frac{N(V_1-V_2)}{2}\xi +\frac{N(V_1+V_2-2U)}{4}\xi^2 \nonumber \\
&& +2N(T^+ +T^- \xi)\sqrt{1-\xi^2}\cos{\theta} \,.
\label{eq:ne10}
\end{eqnarray}
In the case of equal intra-component interaction strengths, i.e., $V_1=V_2$, the Hamiltonian (\ref{eq:ne10}) can be simplified as:
\begin{equation}
H_c=c\xi^2-d\cos{\theta},
\label{eq:ne11}
\end{equation}
with
\begin{eqnarray}
c &=& N\frac{V_1+V_2-2U}{4}\,,
\label{eq:ne12}\\
d &=& -2N(T^+ +T^- \xi)\sqrt{1-\xi^2}\,.
\label{eq:ne13}
\end{eqnarray}
Equation (\ref{eq:ne11}) is analogous to the Hamiltonian of a nonlinear pendulum in which $\xi$ and $\theta$ represents the pendulum's angular momentum and tilt angle, respectively. Here, $d$ can be regarded as the effective length of the pendulum. Importantly, both the magnitude and the sign of $d$ depends on $\xi$. The nonlinear nature manifests itself in the fact that $d$ is not a fixed constant.

At this point, it is instructive to compare our model with another widely studied model, the double well Bose-Hubbard model (DW-BHM) \cite{Shenoy} which describes an atomic condensate confined in a double-well potential. The Hamiltonian of the DW-BHM can be written as:
\begin{equation}
\mathcal{H}^{\rm DW}=\mathcal{E}_l b^{\dagger}_l b_l +\mathcal{E}_r b^{\dagger}_r b_r +\frac{V_l}{2}b^{\dagger2}_l b^2_l+\frac{V_r}{2}b^{\dagger2}_r b^2_r -t(b^{\dagger}_l b_r +b^{\dagger}_r b_l)\,.
\label{eq:ne14}
\end{equation}
Here sub-indicies $l$ and $r$ represent the left and right well, respectively. ${\cal E}_{l,r}$ are bare potential energy of the wells, $V_{l,r}$ are Kerr nonlinear interaction coefficients, and $t$ is the tunneling rate between the well. Following a similar mean-field treatment, we obtain the corresponding semiclassical Hamiltonian as:
\begin{eqnarray}
H^{\rm DW}_c &=& \left(\mathcal{E}_l-\mathcal{E}_r+N\frac{V_l-V_r}{2} \right)\xi +N\frac{V_l+V_r}{4}\xi^2 \nonumber \\
&&-2t\sqrt{1-\xi^2}\cos{\theta} \,,
\end{eqnarray}
which can be simplified under the symmetric double well cases ($\mathcal{E}_l=\mathcal{E}_r$ and $V_l=V_r$):
\begin{equation}
H^{\rm DW}_c=N\frac{V_l+V_r}{4}\xi^2-2t\sqrt{1-\xi^2}\cos{\theta}\,.
\label{eq:ne15}
\end{equation}
This can also be regarded as a Hamiltonian for a nonlinear pendulum, where the effective pendulum length is given by $d^{\rm DW}=2t\sqrt{1-\xi^2}$.

\begin{figure}[h]
\centering
\includegraphics[width=0.9\columnwidth]{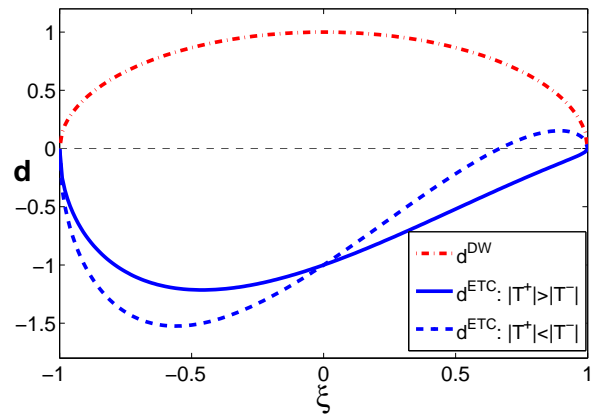}
\caption{(Colour online) The schematic diagram for the odd radius of ETC-BHM. The red-solid line depicts the effective pendulum length of DW-BHM versus the angular momentum $\xi$. The blue-solid (blue-dashed) line depicts the relationship between the effective pendulum length and the angular momentum for the ETC-BHM when $\mid T^+\mid >\mid T^-\mid$ ($\mid T^+\mid <\mid T^-\mid$).}
\label{length}
\end{figure}

Compare the two sets of equations, we can see that the key difference between the two models lies in the nonlinear tunneling term in Hamiltonian (\ref{eq:ne6}) for our model, which is absent in the DW-BHM. As a consequence, in the nonlinear pendulum analog for ETC-BHM, the effective pendulum length $d$ can have either signs, whereas $d^{\rm DW}$ for the DW-BHM is always non-negative. This situation is illustrated in Fig.~\ref{length}. {The positive effective pendulum length implies that the stable equilibrium point is at $\theta=0$ (modulo $2\pi$); whereas for negative effective pendulum length, stable equilibrium occurs at $\theta=\pi$ (modulo $2\pi$). As we will show later, in our model, $d$ can change sign as parameters are tuned, which induces a first-order phase transition when the ground state switches between $\theta =0$ and $\pi$.}

\section{Dynamical Properties}
To illustrate the dynamical properties of the system, we will focus on the case with $V_1=V_2$ and again use the pendulum analog and introduce two types of fundamental dynamical modes: `oscillation' mode and `trapping' mode. Denote $d_x=d \sin\theta$ and $d_y=d \cos\theta$. Thus, within a single dynamical period, we investigate the trajectory of the pendulum bob around the axis in the $d_x$-$d_y$ plane. When the winding number of the trajectory around the axis is zero, the mode is defined as the oscillation mode. While the trajectories with winding number $\pm 1$ correspond to the trapping modes. Two kinds of canonical modes are Josephson Oscillation  (JO) and Self-Trapping (ST), which exist even when only the linear tunneling terms are present. The JO amounts to pendulum bob's vibrating along an arc-shaped segment which goes through some dynamical equilibrium point. The ST corresponds to the case in which an initial angular momentum sufficiently large such that the pendulum bob reaches the top position and continues to rotate with a nonvanishing angular momentum.

\begin{figure}[h]
\centering
\includegraphics[width=3.5in]{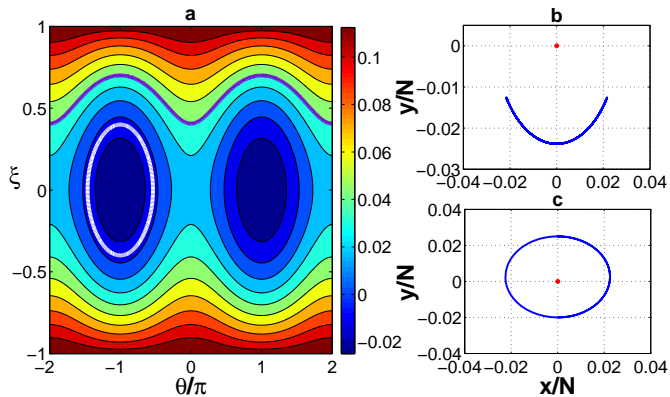}
\caption{(Colour online) (a) The energy distribution of ETC-BHM, {$\frac{H_c}{NU}$, when $T^-=0$, $V_1=V_2=\frac{5U}{4}$, and $T^+=\frac{V_1+V_2-2U}{40}$}; (b) and (c) are the corresponding effective pendulum trajectories of the white contour and the purple one in (a) respectively.}
\label{phase1}
\end{figure}

We now focus on the effects of the nonlinear tunneling. For simplicity, we fix the value of $T^+$ to some positive constant. Under a fixed $T^+$ and with $V_1=V_2$, Hamiltonian (\ref{eq:ne10}) is invariant when $T^- \rightarrow -T^-$ and $\xi \rightarrow -\xi$. Hence we just focus on the case with negative $T^-$. We consider two cases: (1) weak nonlinearity with $|T^-| \le T^+$ and (2) strong nonlinearity with $|T^-|>T^+$.

\subsection{Weak nonlinear case}
We first consider the weak nonlinear case with $|T^-| \le T^+$.

At $T^-=0$, the effective pendulum length is $d=-2NT^+\sqrt{1-\xi^2}$ and the ETC-BHM reduces to the DW-BHM. In this situation, the energy contour lines are depicted in Fig.~\ref{phase1}(a). There are two kinds of dynamical modes: JO and ST. For JO, the energy contour line forms a closed loop in the $\xi$-$\theta$ plane, {and the population difference $\xi$ can change its sign}; while for ST, the energy contour line is an open line for which the sign of $\xi$ remains unchanged. The corresponding   trajectories of the pendulum in the $d_x$-$d_y$ plane are shown in Fig.~\ref{phase1}(b) and (c). In the absence of the nonlinear tunneling, the energy of the pendulum $H_c$ is an even function of both the angular momentum $\xi$ and the tilt angle $\theta$, as can be seen from Fig.~\ref{phase1}(a). Under our choice of $T^+>0$, the effective pendulum length is negative ($d<0$). In the energy contour plot, the center of the JO modes are located at $\xi=0$ and $\theta=\pm\pi$.

With the addition of a nonlinear tunneling strength, the symmetry of the energy $H_c$ about $\xi$ is broken. Fig.~\ref{phase2}(a) and (b) show the the energy contour lines for a weak nonlinear tunneling strength with $T^+\geqslant-T^->0$. When $\xi>0$, the magnitude of the effective length $d$ will be reduced rapidly as the angular momentum $\xi$ grows. This causes the energy of the pendulum to have a weak dependence upon the angle $\theta$. On the contrary, when $\xi<0$, as the magnitude of the angular momentum $\xi$ grows rapidly as the length $d$ is reduced. Therefore, the energy contour lines of the JO and the ST become flat (steep) when $\xi>0$ ($\xi<0$). In Fig.~\ref{phase2}(a) and (b), the orthocenter of the closed loops has been moved down from $\xi=0$.

\subsection{Strong nonlinear case}
\begin{figure*}
\includegraphics[width=6in]{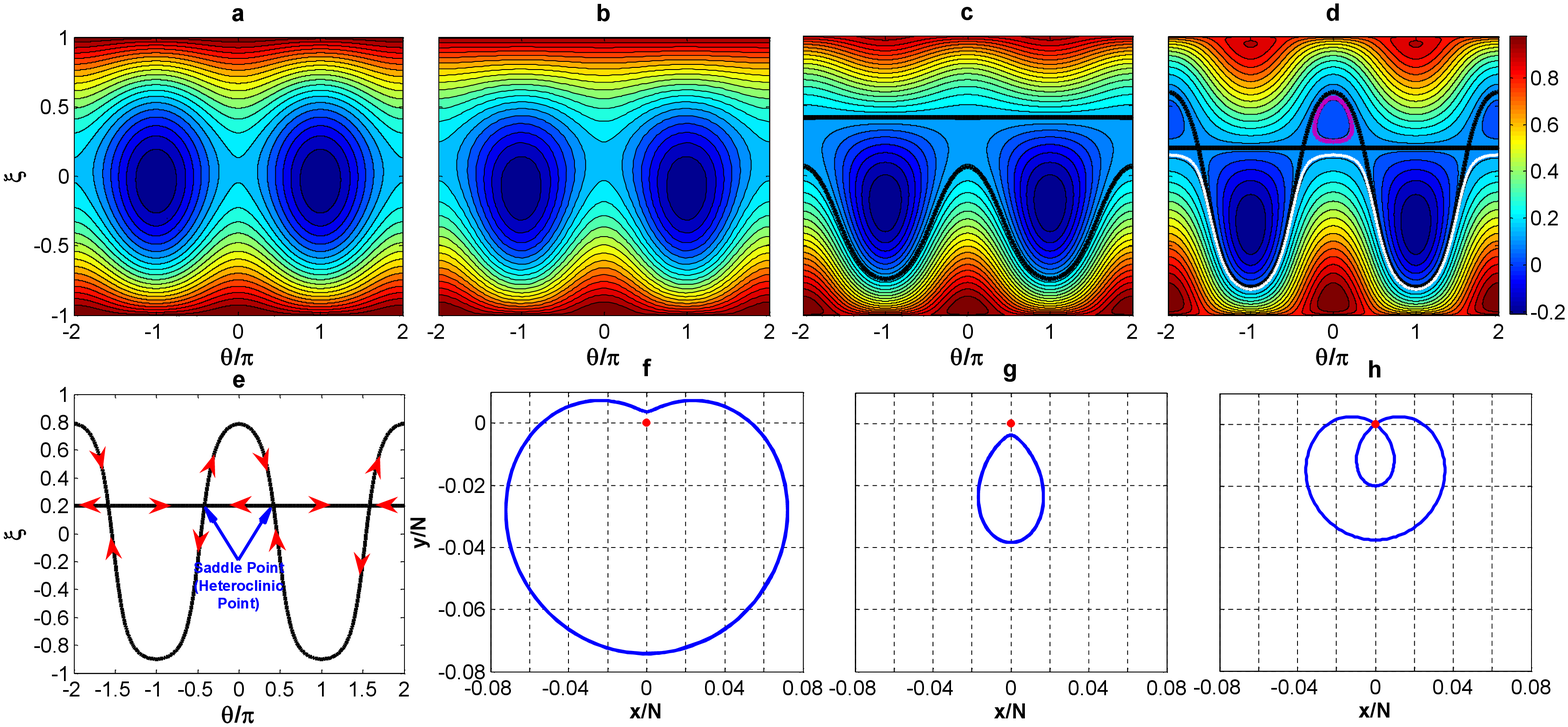}
\caption{(Colour online) Figure (a)-(d) Energy contour plots of ETC-BHM, $\frac{H_c}{NU}$. In all plots, we take $V_1=V_2=\frac{5U}{4}$ and fix $T^+=\frac{V_1+V_2-2U}{40}$. (a) $T^-=-2T^+/5$, (b) $T^-=-T^+$, (c) $T^-=-12T^+/5$, (d)-(h) $T^-=-5T^+$. Figures (f), (g) and (h) depict the trajectories of the effective pendulum length for the white, purple, thick black contour line of energy in Figure (d).}
\label{phase2}
\end{figure*}

We now turn to the strong nonlinear case with
$|T^-|>T^+$. As the nonlinear tunneling strength increases in magnitude, the asymmetry in energy contour about $\xi=0$ becomes more and more dramatic (see Fig.~\ref{phase2}(c) and (d)). There exist closed loops in which the sign of the angular momentum $\xi$ remains positive {(the purple contour in Fig.\ref{phase2}(d))}. Such a mode exhibits characters of both JO (closed loop energy contour) and ST (sign of $\xi$ fixed). From the trajectory of the pendulum bob which is depicted in Fig.~\ref{phase2}(g), one can find that the winding number of the trajectory around the axis is zero. {Therefore, on the one hand, it is a kind of the JO mode based on the definition by winding number; on the other hand, it also has the feature of the ST with the sign of $\xi$ fixed.} Hence, we name this kind of mode Self-Trapping Oscillation (STO). On the other side, some trapping modes with open energy contour lines inherits characters of the JO mode with the sign of $\xi$ varies with $\theta$ {(the white contour in Fig.\ref{phase2}(d))}. The corresponding winding number of the trajectory around the axis is $\pm 1$ (see Fig.~\ref{phase2}(f)). We name this kind of mode the oscillating-type trapping (OTT). The existence of STO and OTT is a distinct feature of the strong nonlinear tunneling effect.

For the strong nonlinear case, there emerges an energy contour line which is just a horizontal line, as can be seen in Fig.~\ref{phase2}(c) and (d). This line corresponds to $\xi=-T^+/T^-$ with the corresponding $\theta$-independent energy $E=H_c(\xi=-T^+/T^-,\theta)=c(T^+/T^-)^2$. However, for this same energy, one can find another energy contour line determined by $H_c(\xi, \theta)=E$ or $c(\xi-\frac{T^+}{T^-})+2NT^-\sqrt{1-\xi^2}\cos\theta=0$. This is represented by the wavy thick black line in Fig.~\ref{phase2}(c) and (d). As $|T^+/T^-|$ decreases, these two contour lines eventually intersect with each other. This is illustrated in Fig.~\ref{phase2}(d), and in Fig.~\ref{phase2}(e) we isolated out these two lines. The intersections of these two lines represent the saddle points of the system, and these two intersecting lines thus become the separatrix of the system \cite{Nonlinear}. The direction of the motion along the separatrix are marked by red arrows in Fig.~\ref{phase2}(e). The trajectory of the pendulum length when the system is prepared on the separatrix is depicted in Fig.~\ref{phase2}(h). It looks like the combination of the STO and the OTT trajectories plotted in Fig.~\ref{phase2}(f) and (g). In real situation, however, the system can come very close to, but never touch, the separatrix. In the region enclosed by the two separatrix, the STO mode arises naturally.

\section{The ground state properties}
\begin{figure}[h]
\centering
\includegraphics[width=0.95\columnwidth]{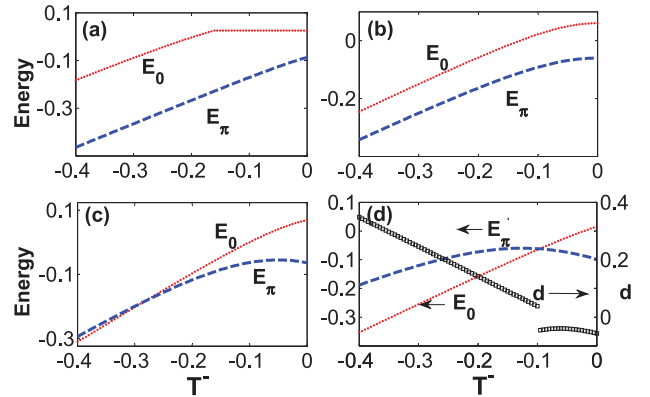}
\caption{(Colour online) Energies $E_0$ and $E_\pi$ (in units of $U$) as functions of $T^-$. The parameters we have used are $T^+=0.03$, $U=1$, $V_1=1.5$, and (a) $V_2=1.2$, (b) $V_2=1.5$, (c) $V_2=1.7$, (d) $V_2=2$. In (d), we also plot the effective pendulum length $d$ for the ground state.}
\label{FOPT}
\end{figure}

We now turn to the study of the ground state. In particular we show that, under proper conditions, there is a first-order phase transition when the nonlinear tunneling rate $T^-$ is tuned. Here we focus on the strong interaction case with $V_{1,2}$, $U \gg \vert T^\pm \vert$. We fix all the parameters except for $T^-$. The mean-field ground state, which minimizes $H_c$ in Eq.~(\ref{eq:ne10}), must occur at $\theta=0$ or $\pi$. We define $E_{0, \pi}={\rm min}\{H_c(\xi, \theta=0, \,\pi)\}/N$ and plot them in Fig.~\ref{FOPT} as functions of $T^-$. As one can see from Fig.~\ref{FOPT}(a) and (b), for $V_1 \geq V_2$, $E_\pi$ remains smaller than $E_0$. However, as shown in Fig.~\ref{FOPT}(c) and (d), when $V_1 < V_2$, a level crossing occurs at a critical value of $T^-$ --- across which the ground state changes from $\theta=0$ to $\theta=\pi$. Hence this critical point represents a first-order phase transition. {We also plot the effective pendulum length $d$ of the ground state in Fig.~\ref{FOPT}(d). We can see that $d$ changes its sign at the critical point. }

This mean-field phase transition can thus be understood as follows. In the strong interaction sitution, the ground state population difference $\xi$ is mainly determined by the first line of Eq.~(\ref{eq:ne10}), which yields: \[\xi_g\approx -\frac{V_1 -V_2}{V_1 +V_2 -2U} \,.\] Under our assumption that $T^+$ and $T^-$ have opposite signs, the coefficent before the $\cos \theta$ term in Eq.~(\ref{eq:ne10}) would change sign when the relative strengths of $T^+$ and $T^-$ are varied as long as $\xi_g >0$, which occurs when $V_1 <V_2$ under our choice of parameters. This leads to the phase transition between $\theta=0$ and $\pi$ in the ground state. {As shown in Fig.~\ref{FOPT}(d), the effective pendulum length $d$ is positive when $\mid T^-\mid$ is larger than the critical value, in this regime, the ground state can be written as $\Psi_0=e^{i\psi}(\sqrt{N_1}, \sqrt{N_2})^T$. On the other side of the critical point, $d$ becomes negative and the ground state reads $\Psi_\pi=e^{i\psi}(\sqrt{N_1}, \sqrt{N_2}e^{i\pi})^T$. Hence this first-order phase transition is a result from the fact that in our model, $d$ is able to change sign as parameters are tuned.} Such a transition would not occur in the DW-BHM when the strength of the tunneling is varied as it would not affect the sign of the $\cos \theta$ term, {and the ground state only takes the form of $\Psi_0$ as the effective pendulum length $d^{\rm DW}$ remains positive}.
%
%

\begin{figure}[h]
\includegraphics[width=\columnwidth]{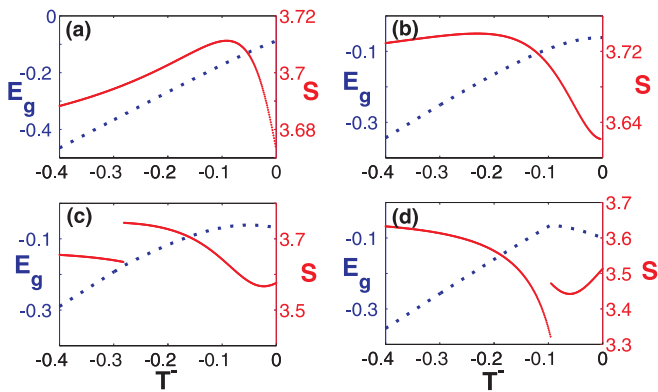}
\caption{(Colour online) The ground state energy $E_g$ (blue dashed lines), in units of $U$, and the von Neumann entropy $S$ (red solid lines) as functions of $T^-$. Here the total number of particles is $N=1000$. The parameters are the same as in Fig.~\ref{FOPT}. }
\label{figED}
\end{figure}

To show that this first-order phase transition is not merely a mean-field artifact, we performed full quantum calculations to find the ground state of ${\cal H}_{\rm eff}$ in Eq.~(\ref{eq:ne6}) via exact diagonalization. In Fig.~\ref{figED} we show the ground state energy $E_g$ and the von Neumann entropy $S$ for the reduced single-particle density matrix as functions of $T^-$. To do so, we expand ${\cal H}_{\rm eff}$ onto the basis $|i, N-i \rangle$ which corresponds to the Fock state with $i$ ($N-i$) particle in the first (second) polariton mode. The ground state has the form \[|\Phi_g \rangle= \sum_{i=0}^N\,c_i |i,N-i \rangle \,,\] from which we can define the von Neumann entropy as \[ S = -\sum_i |c_i|^2 \ln |c_i|^2 \,. \] From Fig.~\ref{figED}, one can see that at the critical value of $T^-$ where the mean-field phase transition occurs, $E_g$ shows a kink and $S$ exhibits a discrete jump. {The discontinuous von Neumann entropy at the critical point is another distinct character of the first order quantum phase transition \cite{EntropyQPT1,EntropyQPT2,EntropyQPT3}.} Hence the quantum calculation confirms the existence of the first-order phase transition.

\section{Detection}
Here we propose an experimental scheme to extract the information on the state of the polaritons. We first map the excitations of the polaritons to the collective atomic excitations ($P^{\dagger}_{1}\doteq S^{\dagger}_{12}$ and $P^{\dagger}_{2}\doteq S^{\dagger}_{13}$) by adiabatically adjusting external laser fields and making the parameter $\Omega$ much smaller than $g$, meanwhile, turning the microwave fields off. In this situation, the effective interactions between the inner energy levels 2 and 3 are frozen. Thus, we may obtain the information on quantum state by probing the population of the atomic energy levels. However, it should be noted that, so far, it is not clear for the relationship between the quantum mechanical eigenstates and mean field kinetic modes. Here, we consider the bosonic condensate quantum mechanical state $\vert \Psi {\rangle}=1/\sqrt{n!}(\cos \zeta S^{\dagger}_{12}+e^{i\eta }\sin \zeta S^{\dagger}_{13})^n\vert vac {\rangle}$, where $\vert vac {\rangle}=\otimes ^{n}_{i=1}\vert 1_i {\rangle}$. The strength of the resonance fluorescence in atomic level 2 is proportional to $\langle N_1 {\rangle}=\sum ^n_{i=0}(n-i)C^i_n(\cos \zeta)^{2(n-i)}(\sin \zeta)^{2i}$. To detect the relative phase $\eta$, we may take advantage of the interference technique, which can be realized by using Hamiltonian: $H_R=\Gamma (S_{13}S^{\dagger}_{12}+S_{12}S^{\dagger}_{13})$. In principle, $H_R$ can be implemented by the optical Raman process. Thus, when the state of the system undergoes the unitary transformation $U(t)=\exp{(-iH_Rt)}$, we may obtain the interference curve dependent on the evolution time by probing the population at the atomic energy level 2. The strength of the resonance fluorescence in atomic level 2 is proportional to $\langle N_1(t) {\rangle}=\sum ^n_{i=0}(n-i)C^i_n[\cos^2 (\zeta-\Gamma t)-1/2 \sin 2\zeta \sin 2\Gamma t(1-\sin \eta)]^{n-i}[\sin^2 (\zeta-\Gamma t)+1/2 \sin 2\zeta \sin 2\Gamma t(1-\sin \eta)]^{i}$. Once the parameters $\zeta$ and $\Gamma t$ are fixed, we may deduce the information on the phase $\eta$.

\section{Conclusion}
In summary, we have made an experimental proposal to realize an extended two-component Bose-Hubbard model in the form of a two-mode polariton condensate inside an optical microcavity. In contrast to the conventional two-component Bose-Hubbard model (such as realized with condensate confined in double-well potential), ours features a nonlinear tunneling term that depends on the number difference between the two modes. We have shown in this work that the nonlinear tunneling term leads to new dynamical modes and induces a first-order phase transition in the ground state of the system.

\begin{acknowledgments}
This work was funded by National Natural Science Foundation of
China (Grant No. 11174270), National Basic
Research Program of China 2011CB921204, 2011CBA00200, Fund of CAS and Research
Fund for the Doctoral Program of Higher Education of China (Grant
No. 20103402110024). Z. -W. Zhou gratefully acknowledges the
support of the K. C. Wong Education Foundation, Hong Kong. H.P. is supported by the Welch foundation (C-1669, C-1681),
and the NSF.
\end{acknowledgments}

\appendix

\section{Derivation of the two-mode Hamiltonian}
\label{app1}
Without loss of generality, here, we set $\omega_1$ as zero point of the energy. Define collective atomic operators:
\begin{equation}
S_{1i}=\frac{1}{\sqrt{N_a}}\sum_{j=1}^{N_a}\vert 1\rangle_{jj}\langle i\vert,
\label{eq:ne2}
\end{equation}
where $i=2,3,...,7$. $S^{\dagger}_{1i}$ refers to the conjugate operator of $S_{1i}$. Under the low excitation limit (LEL) that the number of the excited atoms is far less than the total number of the atoms in the ensemble collective atomic operators approximatively satisfy the relation: $S^{\dagger}_{1i}S_{1k}=\sum_{j=1}^{N_a}\vert i\rangle_{jj}\langle k\vert$, which further leads to the bosonic commutation relation: $[S_{1i}\ ,\ S^{\dagger}_{1j}]=\delta_{ij}$, $[S_{1i}\ ,\ S_{1j}]=0$, and $[S^{\dagger}_{1i}\ ,\ S^{\dagger}_{1j}]=0$. We divide the total Hamiltonian into two parts: $H=H_0+H_I$, $H_0=(\omega_c-\omega_{L1})\sum_{j=1}^{N_a}\vert 2{\rangle}_{jj}\langle 2\vert +(\omega_c-\omega_{L2})\sum_{j=1}^{N_a}\vert 3\rangle_{jj}\langle 3\vert +\omega_c\sum_{j=1}^{N_a}\vert 4\rangle_{jj}\langle 4\vert
+\omega_5\sum_{j=1}^{N_a}\vert 5\rangle_{jj}\langle 5\vert +\omega_6\sum_{j=1}^{N_a}\vert 6\rangle_{jj}\langle 6\vert +(\omega_7-\Delta)\sum_{j=1}^{N_a}\vert 7\rangle_{jj}\langle 7\vert +\omega_c(a^{\dagger}a+\frac{1}{2})$, where $\Delta$ is the detuning of microwave field. Thus, In the rotating frame of $H_0$,  $H^{rot}_I=e^{iH_0 t}H_I e^{-iH_0 t}$ has the following form:
\begin{widetext}
\begin{equation}
\begin{split}
H^{rot}_I&=H^{rot}_{I1}+H^{rot}_{I2}
\\H^{rot}_{I1}&=\varepsilon S^{\dagger}_{12}S_{12} -\varepsilon S^{\dagger}_{13}S_{13} +\delta S^{\dagger}_{14}S_{14} +\Delta S^{\dagger}_{17}S_{17}
+\left(\frac{\Omega_{24}}{2}S^{\dagger}_{12}S_{14}+\frac{\Omega_{34}}{2}S^{\dagger}_{13}S_{14} +\sqrt{N}g_{14}S_{14}a^{\dagger} +h.c.\right)
\\&\quad+\left(\frac{\Omega_{\nu1}}{2}S^{\dagger}_{15}S_{17}+\frac{\Omega_{\nu2}}{2}S^{\dagger}_{16}S_{17}+h.c. \right)
\\H^{rot}_{I2}&=g_{25}e^{-i(\Delta_5+\varepsilon)t}S^{\dagger}_{12}S_{15}a^{\dagger}+g_{36}e^{-i(\Delta_6-\varepsilon)t}S^{\dagger}_{13}S_{16}a^{\dagger} +h.c.
\end{split}
\label{eq:ne3}
\end{equation}
\end{widetext}
Here, $\varepsilon=\omega_2-(\omega_c-\omega_{L1})=(\omega_c-\omega_{L2})-\omega_3$, $\delta=\omega_4-\omega_c$, $\Delta_5=\omega_5-\omega_2-\omega_c$, and $\Delta_6=\omega_6-\omega_3-\omega_c$. Under LEL, $H^{rot}_{I1}$ can be seen as the bosonic quadric form, in which seven bosonic modes can be divided into two independent sub-classes:
$C_1=\{S_{12},S^{\dagger}_{12},S_{13},S^{\dagger}_{13},S_{14},S^{\dagger}_{14},a,a^{\dagger}\}$ and $C_2=\{S_{15},S^{\dagger}_{15},S_{16},S^{\dagger}_{16},S_{17},S^{\dagger}_{17}\}$. Thus, $H^{rot}_{I1}$ can be unitarily diagonalized as:
\begin{equation}
\begin{split}
H^{rot}_{I1}&=\sum_{i=1}^4 \lambda_i P^{\dagger}_i P_i+\sum_{\alpha=0,+,-} \gamma_\alpha Q^{\dagger}_\alpha Q_\alpha.
\end{split}
\label{eq:ne4}
\end{equation}
Here, $P^{\dagger}_i, P_i$ ($Q^{\dagger}_\alpha, Q_\alpha$) are bosonic creation and  annihilation operators generated from the sub-class $C_1$ ($C_2$). $P^{\dagger}_i$ is defined as the polariton mode which is the superposition of cavity mode and collective atomic excitation mode. For simplicity, here, we set $\Omega_{24}=\Omega_{34}=\sqrt{2}\Omega \gg \varepsilon$, $g=\sqrt{N_a}g_{14}$ and $\Delta=0$ to fix the eigenvalues in the $H^{rot}_{I1}$: $\lambda_{1,2}=\pm(\frac{\varepsilon g}{\sqrt{\Omega^2 +g^2 }}+\frac{\varepsilon^3 \Omega^2}{2g(\Omega^2 +g^2)^{3/2}})+\frac{\varepsilon^2 \delta \Omega^2}{2(\Omega^2 +g^2)^2}$,
$\lambda_{3,4}=\frac{\delta \pm \sqrt{\delta^2+4(\Omega^2 +g^2)}}{2} \pm \frac{\varepsilon^2 \Omega^2(\delta^2+2(\Omega^2 +g^2) \mp \delta \sqrt{\delta^2+4(\Omega^2 +g^2)})}{2\sqrt{\delta^2 +4(\Omega^2 +g^2)} (\Omega^2 +g^2)^2}$, $\gamma_0=0$, $\gamma_\pm=\pm \frac{\sqrt{\Omega^2_{\nu1} +\Omega^2_{\nu2}}}{2}$. The condition $\Omega,\delta \gg \varepsilon$ leads to $\vert \lambda_1\vert,\vert \lambda_2\vert \ll \vert \lambda_3\vert, \vert \lambda_4\vert$. Therefore, we name the modes $P^{\dagger}_1$ and $P^{\dagger}_2$ the quasi-dark-state polaritons which can be approximately written as: $P^{\dagger}_{1(2)}\doteq \frac{1}{2}(\frac{g}{\omega}\pm 1)S^{\dagger}_{12}+\frac{1}{2}(\frac{g}{\omega} \mp 1)S^{\dagger}_{13}-\frac{\Omega}{\sqrt{2}\omega}a^{\dagger}$, where $\omega =\sqrt{g^2 +\Omega^2}$.

$H^{rot}_{I2}$ induces the interactions between the bosonic modes in $C_1$ and $C_2$. When large detuning conditions $g_{25} \ll \Delta_5+\varepsilon$ and ${g_{36}} \ll \Delta_6-\varepsilon$ hold, the direct couplings between the modes in different sub-classes are forbidden. However, under appropriate conditions, one can make the second order effect of $H^{rot}_{I2}$ lead to the nonlinear interactions between the interior modes in the sub-class $C_1$ ($C_2$).
Here, we neglect the tedious steps and directly present the approximate conditions to realize the aforementioned idea:
\begin{equation}
\begin{split}
\vert g_m\vert \ll (&\vert \Delta_n\pm \varepsilon\vert,\vert \gamma_\pm \vert,\vert \lambda_1\vert,\vert \lambda_2\vert,\vert \lambda_1-\lambda_2\vert,\vert \gamma_\pm \pm(\lambda_2-\lambda_1)\vert, \\
&\vert \gamma_\pm \pm2(\lambda_2-\lambda_1)\vert) \ll \vert \lambda_3\vert, \vert \lambda_4\vert, \vert \lambda_3 +\lambda_4\vert,
\end{split}
\label{eq:ne5}
\end{equation}
where $m=25,36$ and $n=5,6$. The above first inequality makes the couplings between the bosonic modes in $C_1$ and $C_2$ negligible. The second one makes that the polariton modes 3 and 4 decouple with the other modes under the second order perturbation approximation. Furthermore, assume that vacuum occupation holds for modes $Q^{\dagger}_\alpha (\alpha=0,+,-)$. Under the special resonant condition (see Fig. 2(b)), we can obtain the effective Hamiltonian just including the polariton modes 1 and 2 by the method in \cite{effectiveH}:
\begin{equation}
\begin{split}
\mathcal{H}_{\rm eff}&\approx \frac{V_1}{2}P^{\dagger2}_1 P^2_1 +\frac{V_2}{2}P^{\dagger2}_2 P^2_2 +UP^{\dagger}_1 P_1 P^{\dagger}_2 P_2\\
&\quad +T^+[P^{\dagger}_1(P^{\dagger}_1 P_1 +P^{\dagger}_2 P_2)P_2 +P^{\dagger}_2(P^{\dagger}_1 P_1 +P^{\dagger}_2 P_2)P_1]\\
&\quad +T^-[P^{\dagger}_1(P^{\dagger}_1 P_1 -P^{\dagger}_2 P_2)P_2 +P^{\dagger}_2(P^{\dagger}_1 P_1 -P^{\dagger}_2 P_2)P_1].
\end{split}
\label{eq:appendix}
\end{equation}
which is the effective two-mode Hamiltonian (\ref{eq:ne6}) given that $N=\langle P^{\dagger}_1 P_1 +P^{\dagger}_2 P_2\rangle$. Here the coefficients are given by
\begin{widetext}
\begin{eqnarray*}
V_1 &=& -\frac{\Omega^2}{4\omega^2}\left[g^2_{25}\left(\frac{g}{\omega}+1 \right)^2 \left(\frac{\Omega^2_{\nu2}/\omega^2_{\nu}}{\delta' +2\lambda_2 -2\lambda_1} +\frac{\Omega^2_{\nu1}/(2\omega^2_{\nu})}{\delta' +2\lambda_2 -2\lambda_1 +\gamma_+}+\frac{\Omega^2_{\nu1}/(2\omega^2_{\nu})}{\delta' +2\lambda_2 -2\lambda_1 +\gamma_-} \right)\right. \\
&&\left. +g^2_{36}\left(\frac{g}{\omega}-1 \right)^2 \left(\frac{\Omega^2_{\nu1}/ \omega^2_{\nu}}{\delta' +\lambda_2 -\lambda_1} +\frac{\Omega^2_{\nu2}/ (2\omega^2_{\nu})}{\delta' +\lambda_2 -\lambda_1 +\gamma_+} +\frac{\Omega^2_{\nu2}/ (2\omega^2_{\nu})}{\delta' +\lambda_2 -\lambda_1 +\gamma_-} \right) \right] \,,\\
V_2 &=&-\frac{\Omega^2}{4\omega^2}\left[g^2_{25}\left(\frac{g}{\omega}-1 \right)^2 \left(\frac{\Omega^2_{\nu2}/\omega^2_{\nu}}{\delta'} +\frac{\Omega^2_{\nu1}/(2\omega^2_{\nu})}{\delta' +\gamma_+}+\frac{\Omega^2_{\nu1}/(2\omega^2_{\nu})}{\delta' +\gamma_-} \right) \right. \\
&& \left. +g^2_{36}\left(\frac{g}{\omega}+1 \right)^2 \left(\frac{\Omega^2_{\nu1}/ \omega^2_{\nu}}{\delta' +\lambda_1 -\lambda_2} +\frac{\Omega^2_{\nu2}/ (2\omega^2_{\nu})}{\delta' +\lambda_1 -\lambda_2 +\gamma_+}+\frac{\Omega^2_{\nu2}/ (2\omega^2_{\nu})}{\delta' +\lambda_1 -\lambda_2 +\gamma_-} \right) \right]\,,\\
U &=& -\frac{g^2\Omega^2}{2\omega^4} \left[g^2_{25} \left(\frac{\Omega^2_{\nu2}/\omega^2_{\nu}}{\delta' +\lambda_2 -\lambda_1} +\frac{\Omega^2_{\nu1}/(2\omega^2_{\nu})}{\delta' +\lambda_2 -\lambda_1 +\gamma_+} +\frac{\Omega^2_{\nu1}/(2\omega^2_{\nu})}{\delta' +\lambda_2 -\lambda_1 +\gamma_-} \right) \right. \\
&& \left. +g^2_{36}  \left(\frac{\Omega^2_{\nu1}/ \omega^2_{\nu}}{\delta'} +\frac{\Omega^2_{\nu2}/ (2\omega^2_{\nu})}{\delta' +\gamma_+}+\frac{\Omega^2_{\nu2}/ (2\omega^2_{\nu})}{\delta' +\gamma_-} \right)\right]\,,\\
T^{\pm}&=& -\frac{g g_{25} g_{36}\Omega^2 \Omega_{\nu1}\Omega_{\nu2}}{8\omega^3 \omega^2_{\nu}} \left(\frac{g}{\omega}-1\right) \left[ \left(\frac{1/2}{\delta' +\lambda_2 -\lambda_1 +\gamma_+} +\frac{1/2}{\delta' +\lambda_2 -\lambda_1 +\gamma_-} -\frac{1}{\delta' +\lambda_2 -\lambda_1} \right) \right. \\
&& \left. \pm \left(\frac{1/2}{\delta' +\gamma_+}+\frac{1/2}{\delta' +\gamma_-}-\frac{1}{\delta'} \right) \right]\,,
\end{eqnarray*}
where $\omega_{\nu}=\sqrt{\Omega^2_{\nu1}+\Omega^2_{\nu2}}$. All these coefficients are highly tunable.
\end{widetext}

\begin{figure}[h]
\centering
\includegraphics[width=0.9\columnwidth]{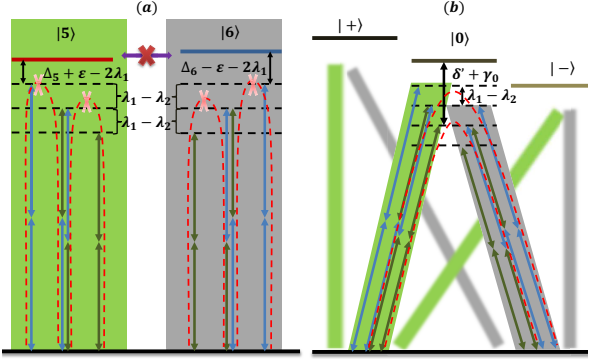}
\caption{(Colour online) Schematic on fourth order transition process of polariton modes 1 and 2. (a) When the microwave fields $\Omega_{\nu1},\Omega_{\nu2}=0$, the fourth-order resonance processes just induce Kerr nonlinearity. Here, the blue (green) arrows refer to $P_1$ and $P^{\dagger}_1$ ($P_2$ and $P^{\dagger}_2$). (b) Three intermediate polariton modes are generated by the resonance process induced by the microwave fields ($\Omega_{\nu1},\Omega_{\nu2}$). On the condition of $\Delta_6 -\varepsilon-(\lambda_1 +\lambda_2)=\Delta_5 +\varepsilon-2\lambda_2=\delta'$, besides the Kerr nonlinear items, $P^{\dagger}_1 P^{\dagger}_1 P_1 P_2$, $P^{\dagger}_2 P^{\dagger}_2 P_2 P_1$ and their conjugate items are induced.}\label{scheme}
\end{figure}

To clarify the physical origin of the terms in Hamiltonian (\ref{eq:appendix}), we present a picture based on the fourth order transition process between polariton modes 1 and 2 as schematically shown in Fig.~\ref{scheme}. When the microwave fields $\Omega_{\nu1}$ and $\Omega_{\nu2}$ are absent, the fourth order resonance terms only include the Kerr nonlinear terms $P^{\dagger2}_i P^2_i (i=1,2)$ and $P^{\dagger}_1 P_1 P^{\dagger}_2 P_2$ as shown in Fig.~\ref{scheme}(a). Here the blue (green) arrows refer to $P_1$ and $P^{\dagger}_1$ ($P_2$ and $P^{\dagger}_2$). With the microwave fields $\Omega_{\nu1}$ and $\Omega_{\nu2}$ present, three more polariton modes $Q^{\dagger}_\alpha (\alpha=0,+,-)$ with eigenfrequency $\gamma_\alpha (\alpha=0,+,-)$ are generated. When the resonant condition $\Delta_6 -\varepsilon-(\lambda_1 +\lambda_2)=\Delta_5 +\varepsilon-2\lambda_2=\delta'$ holds, adiabatically eliminating polariton modes $Q^{\dagger}_\alpha$ leads to terms like $P^{\dagger}_1 P^{\dagger}_1 P_1 P_2$, $P^{\dagger}_2 P^{\dagger}_2 P_2 P_1$ and their Hermitian conjugates. To illustrate this process, here we just single out the intermediate polariton mode $Q^{\dagger}_0$ to illustrate the fourth order resonance transitions (see Fig.~\ref{scheme}(b)). It should be noted that, in this process, the Kerr nonlinear terms are still present. Together with the other similar transition processes, the effective Hamiltonian (\ref{eq:appendix}) is generated.

\begin{figure}[h]
\centering
\includegraphics[width=0.9\columnwidth]{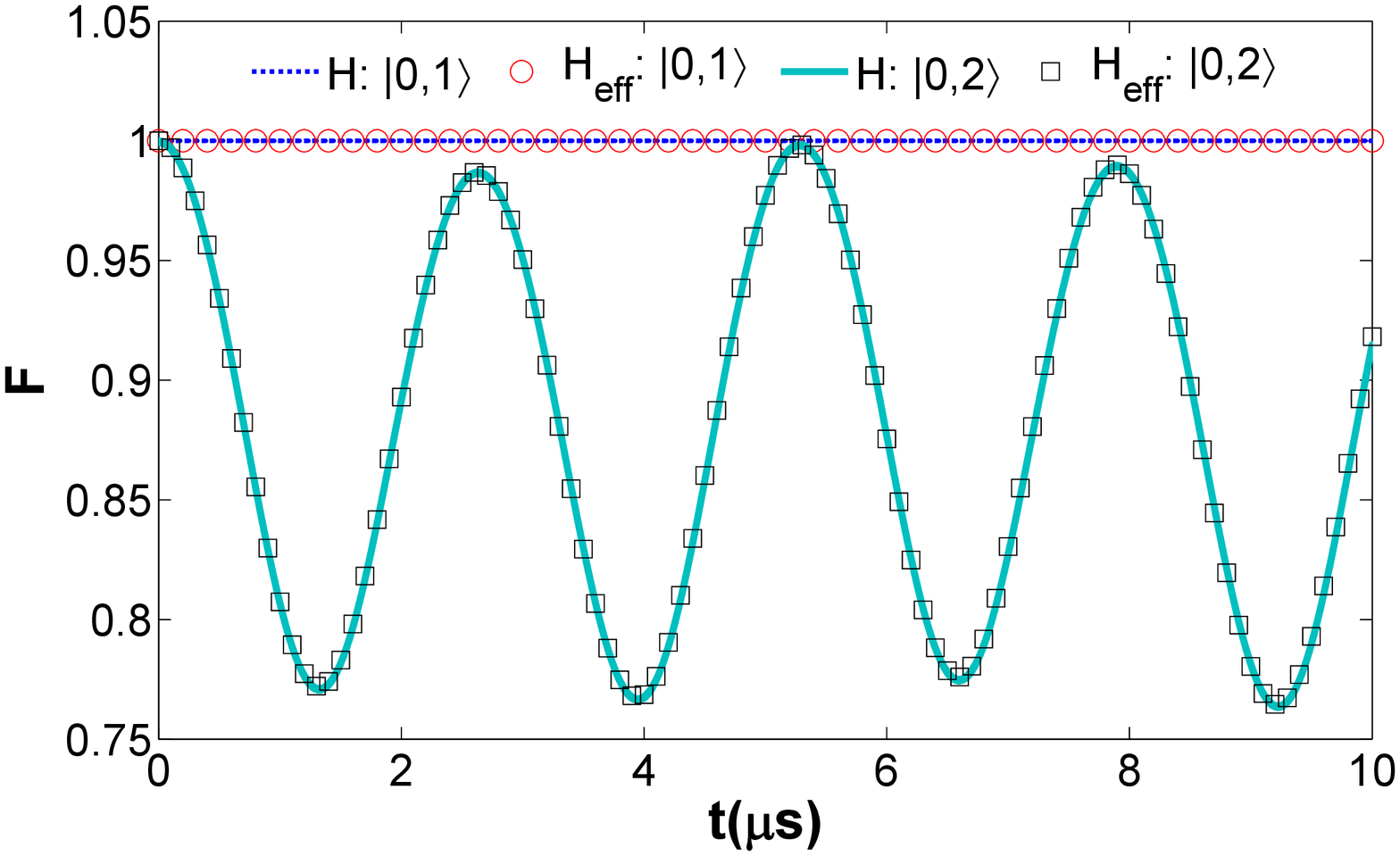}
\caption{(Colour online) The fidelity $F=\vert \langle \phi(t) | \phi(0) \rangle \vert $ versus time. We set Fock state $| 0, 1 \rangle$, and $| 0, 2 \rangle$ as the initial states, where $|N_1,N_2 \rangle$ represents there are $N_1$ polaritons in mode 1 and $N_2$ polaritons in mode 2, and evolve the system under Hamiltonian (\ref{eq:ne3}) (solid lines) or (\ref{eq:appendix}) (dashed lines). }
\label{test}
\end{figure}

To check the validity for approximations invoked in deriving the effective Hamiltonian (\ref{eq:appendix}), we numerically simulate the time evolution in the case of one or two polaritons. As Fig.~\ref{test} shows, the results obtained using Hamiltonian (\ref{eq:ne3}) agrees perfectly with those obtained using (\ref{eq:appendix}).

\begin{thebibliography}{99}

\bibitem{BEC1} M. H. Anderson, J. R. Ensher, M. R. Matthews, C. E. Wieman, and E. A. Cornell, Science, 269, 198 (1995).

\bibitem{BEC2} H.-J. Miesner, D. M. Stamper-Kurn, M. R. Andrews, D. S. Durfee, S. Inouye, and W. Ketterle, Science, 279, 1005 (1998).

\bibitem{BEC3} Anthony J. Leggett, Rev. Mod. Phys. 73, 307 (2001).

\bibitem{BEC4} Markus Greiner, Cindy A. Regal, and Deborah S. Jin, Nature, 426, 537 (2003).

\bibitem{Shenoy} A. Smerzi, S. Fantoni, S. Giovanazzi, and S. R. Shenoy, Phys. Rev. Lett. 79, 4950 (1997).

\bibitem{Ketterle} M. R. Andrews, C. G. Townsend, H.-J. Miesner, D. S. Durfee, D. M. Kurn, and W. Ketterle, Science, 275, 637 (1997).

\bibitem{Oberthaler} M. Albiez, R. Gati, J. Folling, S. Hunsmann, M. Cristiani, and M. K. Oberthaler, Phys. Rev. Lett. 95, 010402 (2005).

\bibitem{Steinhauer} S. Levy, E. Lahoud, I. Shomroni, and J. Steinhauer, Nature, 449, 579 (2007).

\bibitem{Liang} J.-Q. Liang, J.-L. Liu, W.-D. Li, and Z.-J. Li, Phys. Rev. A 79, 033617 (2009).

\bibitem{Fu} H. Cao, and L. B. Fu, Eur. Phys. J. D 66, 97 (2012).

\bibitem{Deng} Hui Deng, Dynamic Condensation of Semiconductor Microcavity Polaritons, PhD thesis, Stanford University, Stanford (2006).

\bibitem{Snoke} R. Balili, V. Hartwell, D. Snoke, L. Pfeiffer, K. West, Science 316, 1007 (2007).

\bibitem{Shelykh} G. Malpuech, D. D. Solnyshkov, H. Ouerdane, M. M. Glazov, and I. Shelykh, Phys. Rev. Lett. 98, 206402 (2007).

\bibitem{Werner} M. J. Werner, A. Imamoglu, Phys. Rev. A 61, 011801 (1999).

\bibitem{Plenio} M. J. Hartmann, F. G. S. L. Brandao, M. B. Plenio, Nat. Phys. 2, 849 (2006).

\bibitem{Liu} An-Chun Ji, Qing Sun, X. C. Xie, and W. M. Liu, Phys. Rev. Lett. 102, 023602 (2009).

\bibitem{Pu} S. K. Adhikari, Hong Lu, and Han Pu, Phys. Rev. A, 80, 063607 (2009).

\bibitem{Nonlinear} John Guckenheimer and Philip Holmes, Nonlinear Oscillations, Dynamical Systems, and Bifurcations of Vector Fields, (Applied Mathematical Sciences Vol. 42), Springer (1985).

\bibitem{EntropyQPT1} L.-A. Wu, M. S. Sarandy, and D. A. Lidar, Phys. Rev. Lett. 93, 250404 (2004).

\bibitem{EntropyQPT2} \"{O}. Legeza, and J. S\'{o}lyom, Phys. Rev. Lett. 96, 116401 (2006).

\bibitem{EntropyQPT3} Huang Hai-Lin, Commun. Theor. Phys. 55, 349 (2011).

\bibitem{effectiveH} Daniel F.V. James, and Jonathan Jerke, Can. J. Phys., 85, 625 (2007).



\end{thebibliography}
\end{document}